# Simultaneous Reduction of Dynamic and Static Power in Scan Structures

Shervin Sharifi, Javid Jaffari, Mohammad Hosseinabady, Ali Afzali-Kusha, and Zainalabedin Navabi
Electrical and Computer Engineering Department
Faculty of Engineering, University of Tehran
Tehran, Iran
{shervin, javid, mohammad}@cad.ece.ut.ac.ir, afzali@ut.ac.ir, navabi@ece.neu.edu

**Abstract**

*Power dissipation during test is a major challenge in testing integrated circuits. Dynamic power has been the dominant part of power dissipation in CMOS circuits, however, in future technologies the static portion of power dissipation will outreach the dynamic portion. This paper proposes an efficient technique to reduce both dynamic and static power dissipation in scan structures. Scan cell outputs which are not on the critical path(s) are multiplexed to fixed values during scan mode. These constant values and primary inputs are selected such that the transitions occurred on non-multiplexed scan cells are suppressed and the leakage current during scan mode is decreased. A method for finding these vectors is also proposed. Effectiveness of this technique is proved by experiments performed on ISCAS89 benchmark circuits.*

## 1. Introduction

Power dissipation during test can be much larger than that of normal operation [1]. However, power constraints are defined for normal mode of operation. Since the current trend is to adopt low-power design techniques and to reduce the package size by exactly matching power dissipation during the circuit normal mode of operation, power constraints defined for the normal operation during the design phase may be much lower than the power consumed during test mode. Therefore, these constraints can be easily exceeded during test causing severe reliability problems. Therefore reducing power dissipation during test application is becoming a critical objective in today's VLSI circuit designs. Moreover, using special cooling equipment to remove excessive heat produced during test application is prevented by the trend toward circuit miniaturization, which makes it especially important to reduce power during test [1].

Power consumption in CMOS circuits can be dynamic or static. Dynamic dissipation occurs as a result of switching activities because of short-circuits current and charging and discharging of load capacitances. Static power consumption is the other portion of the power dissipation in CMOS circuits. Leakage currents including sub-threshold source-to-drain leakage, reverse bias junction band-to-band tunneling, gate oxide tunneling, and other current drawn continuously from the power supply cause static power dissipation.

Leakage current will become an important component in total power consumption because of its exponential relation with decrement in transistor threshold voltage and gate oxide thicknesses that are scaled down in the newer technologies. Since the dynamic power consumption has been the dominant part of power in the older technologies, test power solutions have focused only on this portion of total power dissipation, while static power dissipation is becoming very significant in circuit testing [2].

Scan-based test is the most popular design-for-test (DFT) technique because of its low impact on performance and area. Power problem is a critical issue in this technique. In scan-based test, test vectors are shifted into the scan chains in order to be applied to the circuit-under-test (CUT). Transitions in the scan chain propagate into the CUT and produce several levels of unnecessary switching activities resulting in power dissipation.

Previous works on reducing scan power consumption have tried to reduce switching activity of the circuit.

In this paper, we present a novel solution for power problem in scan-based test. This solution not only reduces the switching activity, but reduces the static power which is consumed in the circuit during test. Hardware area and the required modifications are minimal. Our technique does not have any impact on test time and the maximum working frequency of the circuit in the normal mode. It requires no extra control signals and does not have any routing overhead. Fault coverage is not affected by this method.

The next section reviews some related previous works and Section 3 provides some required backgrounds. The proposed method is explained in Section 4. Section 5 describes and discusses the results and concludes the paper.

## 2. Previous works

Many research works have tried to solve the power problem in scan-based test architectures. [3] proposed an ATPG which exploits all possible "don't cares" that occur during scan shifting, test application, and response capture to minimize switching activity in the circuit under test.

When a scan-based circuit is in the scan mode, transitions of the scan chain may propagate to the combinational part and cause unnecessary power dissipation. In test-per-scan schemes, intermediate values of scan chains do not contribute to the fault coverage. So by preventing scan chain transitions from affecting circuit inputs during scan operations, test power can be reduced without affecting the fault coverage.

[4] proposes a scan chain modification methodology that transforms the stimuli to be inserted to the scan chain through logic gate insertion between scan cells, reducing scan chain transitions.

In [5], scan cells are modified such that scan chain transitions are completely isolated from the combinational part of the circuit.

In enhanced scan structures, hold latches are inserted at scan cell outputs. Before starting a new scan operation, previous values of the scan cells are latched. Therefore, CUT inputs remain unchanged when test data is being shifted in the scan chain.

Reference [6] presents a procedure for inserting test points at the outputs of scan elements of a full-scan circuit in such a manner that the peak power during scan testing is kept below a specified limit while maintaining the original fault coverage. The main drawback of this approach is addition of a global signal to enable test points.

In [7], data required for updating scan vectors are shifted in a separate chain which is included in the design for compression purposes. Scan chain contents are updated after shifting the required data. So the transitions at circuit inputs are limited to the differences between two subsequent scan vectors.

[8] proposes an input control technique to reduce the transition count of the combinational part of a full-scan circuit during test application.

If a pattern can be applied to the PIs during scan such that the propagation through the combinational circuit can be reduced or even eliminated, then the unnecessary power consumption can be saved. An algorithm called C-algorithm is proposed which finds such an input pattern for reducing the number of transitions using a D-algorithm-like method.

As mentioned before, in scaling the transistor physical dimensions and hence the supply voltage due to reliability constraints, the threshold voltage and the gate oxide thickness of the transistors should also be scaled down to keep the drive capability of the transistor and the performance of the digital circuits. This reduction of the threshold voltage and the oxide thickness leads to an exponential increase in the subthreshold and gate leakage current respectively. Several methods have been proposed to address the problem of the static power increase [16]. One of them named input vector control uses a vector which leads to the lowest leakage current and applies it to the primary inputs of the circuit in the standby mode [14], [15].

In [15] an attribute is introduced for each circuit primary input which is called leakage observability. Similar to observability in the area of test pattern generation, the leakage observability indicates the degree to which the value of a particular circuit input is observable in the magnitude of leakage from power supply. In other words, it predicts the average effect on leakage if a primary input is set to a 1 or to a 0. We have extended this attribute to intermediate signals, so it can be used as a directive when justifying transition blocking values on intermediate lines through setting circuit inputs to proper values.

## 3. Background

For a CMOS circuit, total power consists of dynamic and static components at active mode. Dynamic power is proportional to the number of transitions in circuit, and static part of power is due to leakage currents of gates.

### A. Dynamic Power Estimation

Ignoring direct-path short-circuit current, dynamic power dissipation is mostly due to charging and discharging of load and internal capacitances, which can be obtained as follows:

$$P_{Dyn} = P_{Dyno} + P_{Dynj}$$
$$= \frac{1}{2} f \left( V_{DD}^2 \sum_i (\alpha_i C_{Li}) + V_{DD} \sum_i \sum_j (\alpha_{ij} C_{ij} V_{ij}) \right) \quad (1)$$

In the above, $P_{dyno}$ and $P_{dynj}$ are dynamic powers due to the load and internal capacitances, respectively. $f$ is the clock frequency. $i$ represents the gate $i$ and $j$ denotes the $j$th internal node in a gate. The switching activities at gate $i$ and at the $j$th internal node of gate $i$ are represented as $\alpha_i$ and $\alpha_{ij}$, respectively. $V_{ij}$ is the voltage swing of the $j$th internal node of gate $i$, which is equal to $V_{DD}$-$V_{th}$. $C_{Li}$ and $C_{ij}$ are the load capacitance and the $j$th internal node capacitance of gate $i$, respectively.

### B. Static Power Estimation

The total leakage current of a logic gate includes two major components, namely, subthreshold and gate leakage.

The subthreshold leakage current is one of the important components of leakages in CMOS digital circuits. This component exponentially increases with the reduction of the threshold voltage [9]. Using the Berkeley short-channel IGFET (BSIM) MOS transistor model [10], the subthreshold current is approximated as

$$I_{Sub} = A e^{\frac{q}{nkT}(V_{GS} - V_{T0} - \delta V_S + \eta V_{DS})} \left( 1 - e^{\frac{-qV_{DS}}{kT}} \right) \quad (2)$$

Where $kT/q$ is the thermal voltage, $n$ is the subthreshold swing coefficient of the transistor, $V_{DS}$ is the drain to source voltage, $V_{GS}$ is the gate to source voltage, $V_{T0}$ is the zero bias threshold voltage, $\delta$ is the body effect coefficient, $\eta$ is the drain induced barrier lowering (DIBL) coefficient, and

$$A = \mu_0 C_{ox} \frac{W_{eff}}{L_{eff}} \left(\frac{kT}{q}\right)^2 e^{1.8} \qquad (3)$$

Here, $\mu_0$ is the zero bias mobility and $C_{ox}$ is the gate oxide capacitance per unit area. Equation (2) suggests that the subthreshold current for each transistor be estimated when the terminal voltages are known. For the transistors in a logical gate, the terminal voltages depends on the gate input signals as well as the gate topology. The voltages may be easily calculated for a parallel combination of transistors (e.g., the pull-up network of an *n*-input NAND gate) where $V_{DS}$'s are the same for all parallel transistors. This is not the case for the transistors in series (e.g., in the pull-down network of an NAND gate).

The gate leakage is due to the tunneling of an electron (or hole) from the bulk silicon through the gate-oxide potential barrier into the gate [12]. Direct tunneling is modeled as shown below [13].

$$J_{DT} = A\left(\frac{V_{ox}}{T_{ox}}\right)^2 \exp\left(\frac{-B\left(1-\left(1-\frac{V_{ox}}{\phi_{ox}}\right)^{\frac{3}{2}}\right)}{\frac{V_{ox}}{T_{ox}}}\right) \qquad (4)$$

Where $J_{DT}$ is the direct tunneling density, $V_{ox}$ is the drop across the thin oxide, $\phi_{ox}$ is the barrier height for the tunneling particle (electron or hole), and $T_{ox}$ is the oxide thickness. *A* and *B* are physical parameters and can be found in [13] with more details. It can be observed from Equation (4) shows that the tunneling current increases exponentially with a decrease in oxide thickness as well as increase in $V_{ox}$. The latter depends on the biasing condition which is related to the gate topology and input signal. Therefore, the input pattern of each gate strongly affects the subthreshold as well as the gate leakage current.

To avoid complex calculations for estimation of total leakage we have used an HSPICE BSIM4 simulator to obtain total leakage currents for the transistors of the gates with different input signal levels. The results are stored in several tables containing the leakage of each gate for a given input pattern.

Finally, the total leakage power can be expressed as follows [11].

$$P_{Sub} = \sum_i I_{Sub,i} V_{DD} \qquad (5)$$

Where $I_{Sub,i}$ is the standby leakage current through each gate *i*.

### C. Leakage Observability

As mentioned before, [15] introduced an attribute named *leakage observability* which is used for each primary input line. Leakage observability represents a measure of difference between overall leakage costs when a primary input line is set to one or zero. In other words, the magnitude of this attribute for each primary input line indicates how a binary value on that line can influence the total leakage. The leakage observability of input line *i* is defined by (6).

$$L_{obs}(i) = L_{avg}(i,1) - L_{avg}(i,0) \qquad (6)$$

$L_{avg}(i, v)$ is the average leakage cost for input i forced to value v. Calculation of leakage observability in [15] is performed in reverse topological order on a network. Leakage observabilities of all lines are calculated, but only leakage observabilities on input lines are used in [15] to find the minimum leakage pattern.

## 4. The Proposed Method

The proposed method tries to reduce both dynamic and static power dissipation during scan-based testing without affecting performance and with minimal area overhead.

Some previous solutions like [5] and enhanced scan try to reduce dynamic power dissipation during scan by isolating scan chain transitions from the combinational parts of the circuit by adding latches or tri-state buffers to the scan-cell outputs (pseudo-inputs of the combinational part of the circuit). These methods cause performance degradation by adding logic to the outputs of all scan cells.

Our proposed method tries to block some of scan-chain transitions without affecting the performance of the circuit during normal operation. These transitions are blocked by using multiplexers at scan-cell outputs which allow applying desired values while scanning the vectors in the chain. To avoid performance degradation, multiplexing is performed only on those pseudo-inputs that are not on the critical path(s) of the circuit. If primary inputs of the circuit are accessible, any desired value can be applied to the multiplexed pseudo-inputs and primary inputs of the circuit during the scan phase. We call primary inputs and multiplexed inputs as *controlled inputs* of the circuit. Transitions on the non-multiplexed pseudo-inputs can still affect the combinational part hence resulting in unnecessary power dissipation.

The proposed method minimizes or eliminates this power dissipation by trying to suppress these transitions as near as possible to their origin (scan cell outputs). This is performed by applying appropriate patterns to the controlled inputs of the circuit. This method significantly reduces power dissipation during scan operations. The structure of this method for a full-scan sequential circuit is shown in Figure 1. Controlled inputs are shown with dashed lines.

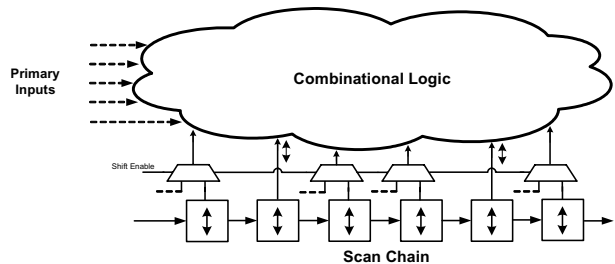

**Figure 1 The proposed method**

As stated before, static power dissipation will be the dominant portion of the total power dissipation in future technologies. The proposed structure is used in order to reduce the static power dissipation. The static power reduction is based on an input vector control technique combined with a method which reorders gate inputs. In the input vector control technique, an input pattern is applied to the circuit inputs which minimize the leakage current. In our structure, this pattern can be applied to the circuit through the set of controlled inputs.

Therefore, in order to reduce both dynamic and static power dissipations, an appropriate pattern should be found to be applied to the controlled inputs during scan mode. When the scan operation finishes and scan values reach their corresponding scan cells, the circuit enters its normal mode of operation. In the normal mode, inserted multiplexers are switched to scan cell outputs and scan cell contents are applied to the circuit pseudo-inputs. The select line of a multiplexer can be connected to the shift enable signal available in all scan structures. In all scan structures, all scan cells receive the shift enable signal. So no extra control signal is required for this method.

An algorithm is also proposed which finds the desired vector. This vector should minimize the leakage current while suppressing the transitions originated from non-multiplexed pseudo-inputs.

The proposed method consists of these major steps.
1. Identifying pseudo-inputs suitable for being multiplexed and adding multiplexers to them (performed by AddMUX() procedure)
2. Finding the appropriate vector for controlled inputs (performed by the procedure called *FindControlledInputPattern()* )

The first step identifies pseudo-inputs which can be multiplexed without affecting the performance. This step is performed as follows:

```
AddMUX()
1.Find delay of critical path(s) of the circuit
2.For each pseudo-input PI
        a.Add a multiplexer to PI
        b.If the critical path delay of the circuit
          has changed after inserting the multiplexer,
          remove the multiplexer
```

First, the critical path delay of the circuit is extracted. Then, the multiplexers of those inputs which affect the critical path delay are removed.

The next step is finding the appropriate pattern for the controlled inputs which minimizes the leakage current and suppresses the scan chain transitions.

An algorithm is proposed to find such an input pattern. The basic idea is that there are many vectors that can disable transitions propagating from non-controlled pseudo-inputs of the circuit. This algorithm is based on a method of finding transition-blocking vectors which is directed by leakage observability. This algorithm is similar to C-Algorithm [8], but is extended and directed by leakage observability.

To describe the proposed procedure *(FindControlledInputPattern())*, some concepts should be defined first.

Based on values assigned to controlled inputs, transitions may propagate to some nodes. These nodes are called *tn* (*transition node*). Set of all transition nodes is called *Transition Node Set* (*TNS*). Each *tn* is connected to input of a gate. Each of these gates is called *tg* and set of all *tg*s is called *Transition Gate Set* (*TGS*).

```
FindControlledInputPattern()
1.Initialize TNS to the set of non-multiplexed pseudo-inputs.
2.Update TNS, TGS
3.Repeat
    a.   Get a gate from TGS with the largest output
         capacitance (mc_tg). The corresponding tn
         is called mc_tn.
    b.   cv = controlling value of mc_tg.
    c.   is_transition_blocked = false;
    d.   Repeat
         i.   Select an input node candidate_input
              of gate mc_tg with don't care value
              (other than mc_tn).
              If there is more than one option, select
              based on leakage observability).
         ii.  is_transition_blocked =
              Justify(candidate_node, cv)
              (Justify() is directed by leakage
              observability)
         iii. If (is_transition_blocked=true) Goto f
    e.   Until all don't care inputs of the
         mc_tg are checked
    f.   Add all fan-out nodes of mc_tg to TNS
    g.   Update TNS, TGS
  Until TGS becomes empty
4.Save the assigned values on controlled inputs
```

After trying each transition gate, *TNS* and *TGS* are updated. Process of updating *TNS* and *TGS* are described here:

```
Update TNS, TGS
1.Repeat
        a.Get a node tn_i from TNS
        b.target_gate= gate connected to the tn_i output
        c.if target_gate is NOT, XOR, XNOR or FANOUT,
          add its output line(s) to TNS, Goto a
        d.if any input of the target_gate has controlling
          value, Goto a
        e.if all inputs of the target_gate have non-controlling
          value, add its output line(s) to TNS
  Until all nodes of TNS are processed
2.For each transition node in updated TNS, put its target gate in TGS
```

As stated before, leakage observability is used to direct the process of finding the input control pattern.

The algorithm should make decisions in different steps to limit the large space of possible solutions. Two important types of decisions are made in this algorithm. The first decision should be made when the algorithm selects which input of a transition gate is to be set to the controlling value. The second type of decision is made in the Justification process. Justification in this algorithm is performed by a PODEM-like method. In each step of this algorithm, one transition point is tried to be suppressed by applying a controlling value to the input(s) of its target gate. So the objective is setting an internal node to the desired value. Mapping this objective to values required at the controlled inputs is performed by the Backtrace procedure. Backtrace starts from the objective node and traverses internal lines toward the controlled inputs. On its way toward the controlled inputs, when reaching a new gate, Backtrace selects one of the gate inputs with a don't care value. Both of these types of decisions are made based on leakage observability.

In [15], leakage observability was used only in the primary input lines in order to find the minimum leakage vector. We have extended the use of leakage observability to all circuit lines in order to direct the *FindControlledInputPattern()* procedure.

According to the definition of leakage observability, larger leakage observability means larger difference between average leakages in '1' and '0' states of a line. Therefore, when deciding on inputs of a gate, if the value to be set is '1' ('0'), we choose the input with minimum (maximum) leakage observability. Using this directive allows us to select a low leakage vector out of all possible vectors which can block the scan chain transitions.

When *FindControlledInputPattern()* finishes its work, there are still some controlled inputs that are not assigned values. These don't care controlled inputs can be used to further reduction of power consumption. A simulation-based method is used to find the minimum-leakage vector for these inputs. The appropriate values for these don't care inputs to reduce the total circuit leakage current can be found by applying several random inputs and examining the total leakage for each of them. The number of the required simulations is far less than the total possible vectors [14].

After finding the appropriate vector for the controlled inputs, they are applied to the circuit to find the values of the internal nodes of the CUT. These values are used as directives to change input of each gate in order to reduce the total gate leakage. As mentioned before the leakage current of a gate is strongly related to the pattern applied to that gate. So in some cases changing the inputs of a gate can be helpful for reducing its leakage.

For example, as can be observed from Figure 1, the leakage current of a NAND2 gate is strongly different in "01" and "10" states. So changing the order of inputs such that it will result in "01" rather than "10" can further decrease the total leakage in scan mode. This method is used globally for the circuit and the best order of inputs is found and applied to the circuit.

| A | B | Leakage (nA) |
|---|---|---|
| 0 | 0 | 78 |
| 0 | 1 | 73 |
| 1 | 0 | 264 |
| 1 | 1 | 408 |

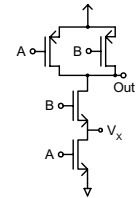

**Figure 2 Leakage current of NAND2 gate in 45nm technology**

## 5. Results and Conclusions

To verify the improvement in power dissipation, the proposed technique was compared with the traditional scan structure; also, the input control structure [8] was implemented and compared with our technique. The proposed method was implemented using C++ language and tested on ISCAS89 benchmarks. A technology mapping was used to map the circuit to a library, which contains only NAND gates, NOR gates, and inverters. SPICE simulation results were obtained for CMOS 45nm technology with the supply of 0.9v. The minimum feature size of 45nm was chosen for the channel length while the widths were selected for the minimum power delay product.

Our experiments were performed using test vectors generated by ATOM [18]. No test vector reordering or scan cell reordering was performed in these experiments. By applying reordering techniques, further improvements can be achieved.

Table I compares dynamic and static parts of power dissipated in the combinational part of the ISCAS89 circuit shown for traditional scan, input control technique and the method proposed here. The values in the dynamic columns must be multiplied by the working frequency to give the actual dynamic power. Static portion of power is not currently as effective as the dynamic one in the total dissipated power, but it will outreach the dynamic portion in future technologies. This table shows a fair amount of reduction in power dissipation especially in the dynamic portion.

The proposed method has reduced the total power dissipation including static and dynamic power while it does not have any impact on test time. It also does not affect the critical path of the circuit, so it does not affect the maximum working frequency of the circuit. It does not incur routing overhead since inputs of the multiplexer can be locally connected to Vcc or Gnd. It also requires no extra control signal since it uses the *Shift Enable* signal as its control signal (This signal is available in all scan cells of the circuit).

This method can be used as an efficient low-overhead solution for power problem in scan-based DFT structures.

**Table I. Power dissipation for our proposed and prior structures**

| Circuit | Traditional Scan Structure | | Input Control [8] | | Proposed Structure(μW) | | Improvement Compared with Traditional Scan (%) | | Improvement Compared With Input Control [8] (%) | |
|---|---|---|---|---|---|---|---|---|---|---|
| | Dynamic (/f) (μW/Hz) | Static (μW) | Dynamic (/f) (μW/Hz) | Static (μW) | Dynamic (/f) (μW/Hz) | Static (μW) | Dynamic | Static | Dynamic | Static |
| s344  | 5.88E-8 | 27.99  | 5.72E-8 | 27.50  | 3.24E-8 | 23.89  | 44.82 | 14.65 | 43.23 | 13.12 |
| s382  | 6.43E-8 | 27.58  | 5.51E-8 | 26.69  | 2.38E-8 | 24.42  | 62.90 | 11.46 | 56.73 | 8.50  |
| s444  | 8.00E-8 | 33.72  | 6.92E-8 | 33.30  | 2.44E-8 | 27.99  | 69.44 | 17.00 | 64.67 | 15.95 |
| s510  | 8.46E-8 | 47.93  | 8.18E-8 | 47.50  | 8.22E-8 | 45.96  | 2.92  | 4.11  | -0.41 | 3.24  |
| s641  | 5.69E-8 | 59.07  | 1.77E-8 | 56.97  | 1.78E-8 | 48.97  | 68.80 | 17.10 | -0.5  | 14.05 |
| s713  | 6.30E-8 | 66.15  | 1.85E-8 | 64.90  | 1.82E-8 | 52.10  | 71.06 | 21.23 | 1.25  | 19.71 |
| s1196 | 3.10E-8 | 115.54 | 3.06E-8 | 117.75 | 2.52E-8 | 95.78  | 18.61 | 17.09 | 17.50 | 18.65 |
| s1238 | 3.19E-8 | 121.56 | 3.39E-8 | 124.75 | 2.59E-8 | 96.38  | 18.64 | 20.70 | 23.63 | 22.74 |
| s1423 | 2.24E-7 | 128.22 | 1.93E-7 | 130.23 | 5.43E-8 | 117    | 75.77 | 9.02  | 71.83 | 10.43 |
| s1494 | 3.56E-7 | 177.52 | 3.48E-7 | 179.86 | 3.52E-7 | 164.87 | 9.52  | 7.12  | 7.45  | 8.33  |
| s5378 | 8.90E-7 | 327.52 | 1.29E-8 | 332.02 | 1.17E-8 | 315    | 98.68 | 3.82  | 9.50  | 5.12  |
| s9234 | 1.50E-6 | 819.98 | 1.68E-8 | 854.52 | 1.57E-8 | 772.36 | 98.95 | 5.80  | 6.96  | 9.61  |

## References

[1] P. Girard, "Survey of Low-Power Testing of VLSI Circuits", IEEE Design & Test of Computers, Vol.19, No.3, May/June 2002.

[2] Kaushik Roy, T.M. Mak, and Kwang-Ting Cheng, "Test Consideration for Nanometer Scale CMOS Circuits", Proceedings of the 21st IEEE VLSI Test Symposium (VTS.03), 2003.

[3] S. Wang, and S. K. Gupta, "An Automatic Test Pattern Generator for Minimizing Switching Activity During Scan Testing Activity", IEEE Transactions on Computer-Aided Design of Integrated Circuits and Systems, Vol. 21, No. 8, August 2002, pp.954-968.

[4] O. Sinanoglu A. Orailoglu, "Modeling Scan Chain Modifications for Scan-in Test Power Minimization", Proceedings of International Test Conference (ITC 2003), 2003, pp. 602 - 611

[5] A. Hertwig, H.J. Wunderlich, "Low Power Serial Built-In Self-Test", European Test Workshop, 1998. pp. 49-53

[6] Ranganathan Sankaralingam, Nur A. Touba, "Inserting Test Points to Control Peak Power During Scan Testing", IEEE International Symposium on Defect and Fault Tolerance in VLSI Systems, 2002.

[7] S. Sharifi, M. Hosseinabadi, P. Riahi, Z. Navabi, "Reducing Test Power, Time and Data Volume in SoC Testing Using Selective Trigger Scan Architecture", IEEE International Symposium on Defect and Fault Tolerance in VLSI Systems, 2003.

[8] T. C. Huang, and K. j. Lee, "Reduction of power consumption in scan-based circuits during test application by an input control technique", IEEE Transaction on Computer-Aided Design, Vol. 20, No. 7, July 2001, pp. 911-917.

[9] K. Roy, S. Mukhopadhyay, and H. Mahmoodi Meymand, "Leakage current mechanisms and leakage reduction technique in deep-submicrometer CMOS circuits, " Proceeding of the IEEE, Vol. 91, No. 2, pp. 305-327, Feb. 2003.

[10] B. J. Sheu, D. L. Scharfetter, P. K. Ko, and M. C. Teng, "BSIM: Berkely short-channel IGFET model for MOS transistors," IEEE Journal of Solid-State Circuits, Vol. SC-22, pp. 558-566, Apr. 1987.

[11] Jan M. Rabaey, Anantha Chandrakasan, and Borivoje Nikolic, Digital Integrated Circuits, Prentice-Hall, 2003.

[12] S. Mukhopadhyay, C. Neau, R. T. Cakici, A. Agarwal, C. H. Kim and K. Roy, "Gate leakage reduction for scaled devices using transistor stacking, "IEEE Transaction on Very Large Scaled Integration (VLSI) Systems. Vol. 11, No. 4, pp. 716-730, Aug. 2003.

[13] K. Schuegraf and C. Hu, "Hole injection SiO2 breakdown model for very low voltage lifetime extrapolation," IEEE Transaction on Electron Devices, Vol. 41, pp. 761-767, May. 1994.

[14] J. P. Halter and F. N. Najm, "A gate-level leakage power reduction method for ultra-low-power CMOS circuits," IEEE Custom Integrated Circuits Conference, 1997.

[15] M. C. Johnson, D. Somasekhar, and K. Roy, "Models and algorithms for bounds on leakage in CMOS circuits, " IEEE Transactions on CAD of Integrated Circuits and Systems, Vol. 18, No. 6, June 1999, pp. 714-725.

[16] K. Roy, S. Mukhopadhyay, and H. Mahmoodi Meymand, "Leakage current mechanisms and leakage reduction technique in deep-submicrometer CMOS circuits, " Proceeding of the IEEE, Vol. 91, No. 2, pp. 305-327, Feb. 2003.

[17] Berkeley Predictive Technology Model, "http://www-device.eecs.berkeley.edu/~ptm".

[18] I. Hamzaoglu and J. H. Patel, "New Techniques for Deterministic Test Pattern Generation," Proc. VLSI Test Symp, pp.446-452, April 1998.